# Design Spaces and How Software Designers Use Them: a sampler

Mary Shaw
School of Computer Science
Carnegie Mellon University
Pittsburgh PA, USA
mary.shaw@cs.cmu.edu

Marian Petre
School of Computing and Communications
Open University
Milton Keynes, UK
m.petre@open.ac.uk

## ABSTRACT

Discussions of software design often refer to using "design spaces" to describe the spectrum of available design alternatives. This supports design thinking in many ways: to capture domain knowledge, to support a wide variety of design activity, to analyze or predict properties of alternatives, to understand interactions and dependencies among design choices. We present a sampling of what designers, especially software designers, mean when they say "design space" and provide examples of the roles their design spaces serve in their design activity. This shows how design spaces can serve designers as lenses to reduce the overall space of possibilities and support systematic design decision making.

## 1. What are design spaces?

As Simon famously said, "Everyone designs who devises courses of action aimed at changing existing situations into preferred ones" [36]. We consider here design spaces, ways of describing existing situations and possible preferred ones, and the ways design spaces are used by software designers. Design spaces are widely used by designers in many domains. We draw on that broad experience to provide context, examples, and guidance for their use in software engineering.

Design spaces embody the design alternatives for problem domains or applications. In practical systems, both the design alternatives and the dependencies among the choices for those alternatives are numerous and open-ended. This creates a dilemma for designers: how to reduce the intellectual complexity of the open-ended "space of possibilities" to something manageable by focusing on a subset of the complete space that is most helpful in the current state of the design, i.e., a "design space". In identifying a design space, a designer chooses some perspective on the "space of possibilities" intended to help reduce the number of alternatives and dependencies, then navigates within that selected subset [4][9][23][28][43]. The different perspectives, like lenses, have a particular "focal range" or orientation (e.g., structured vs integrated, as discussed in the sections that follow) that shapes both the priorities and the process of exploration.

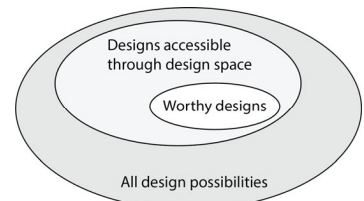

**Figure 1: Distinctions among the space of all possibilities, the subset of designs accessible in a design space, and the designs in that subset that are worthy of consideration. Not to scale.**

Woodbury and Burrow [43] recognize the vastness of the space of all design possibilities, with finite but incomprehensibly large numbers of possibilities, and with worthy designs a vanishingly small subset of the space. In such a setting, the problem is not whether a good design exists but whether it is accessible to a designer searching the space. The full set of design alternatives is usually quite rich and quite interconnected, and a decision about one aspect of the design influences choices about other aspects. Schön says "A designer makes things.… Typically this making process is complex. There are more variables—kinds of possible moves, norms, and interrelationships of these—than can be represented in a finite model" [29]. Hence design spaces are helpful—but incomplete.

Engineering practice includes both routine and innovative (also called normal and radical) design tasks; the former involve familiar problems and reuse of large portions of prior work, and the latter calls for novel solutions to unfamiliar problems [30]. Within routine design the design spaces have arguably been well-mapped for well-scoped domains. Innovative design, by its nature, requires the designer to engage deeply with the alternatives. Hence the present discussion focuses on innovative design—including engagement with "wicked" problems that resist solution [25] —where the choice of design space and the nature of the designer's engagement with design alternatives is instrumental[1].

Designers sometimes address the complexity of the design space by taking a structured approach, making simplifying assumptions to manage the complexity, for example by (provisionally) assuming independence of decisions or focusing on a few principal aspects of the design[2]. At other times, they undertake integrative exploration, keeping the dependencies front-of-mind as they explore their problem understanding and design options[3]. Broadly speaking, structured spaces tend to be associated with explicit knowledge; depth-first search of the design space; treatment of decisions as independent; and reductionist reasoning. On the other hand, integrative design spaces tend to be associated with tacit knowledge; breadth-first search of the design space; attention to dependencies among decisions; and holistic reasoning. In both cases, designers often revise the aspects under consideration as the design evolves[4]. The choice between integrative and structured lenses leads to different approaches to the design space, different representations, and different uses[5].

These lenses are discussed in more detail in the sections that follow, with concrete examples from the literature both in software design and other domains.

## 2. Structured design spaces for domains

Some design processes create explicit, structured descriptions of the design alternatives. This often focuses primary attention on the design choices rather than the dependencies among these choices. They address the complexity of the design space by making simplifying assumptions such as focusing—for the moment—on a few of the most significant design decisions and their alternatives; they may assume independence among the choices[6]. This is comparable to the engineer's inclination to use linear models whenever possible: they're simple, understandable, tractable, and often good enough[7].

This approach leads to design spaces with descriptions such as "The design space in which a designer seeks to solve a problem is the set of decisions to be made about the designed artifact together with the alternative choices for these decisions. … Intuitively, a design space is a discrete Cartesian space in which the dimensions correspond to design decisions, the values on each dimension are the choices for the corresponding decision, and points in the space correspond to complete design solutions" [33]

A key to these examples is creating an explicit representation of the design decisions, the alternative choices, and perhaps the dependencies. These examples show a variety of representations; each is presumably appropriate for its problem and might not be so for other problems[8]. Some are largely Cartesian, some are hierarchical, some are more complex. Some are graphical, some are textual, some are mathematical models. For some the alternatives are ratio-scale measures, for others some of the dimensions are on nominal or cardinal scales.

Design spaces most often address the problem or requirement space (what the client needs) and the solution or implementation space (how the implementation will accomplish that). They may emphasize functionality, quality attributes, or value information.

The vastness of the design space arises from the open-ended set of possible design decisions. Not only does this lead to combinatorial explosion, it flies in the face of conventional assumptions that specifications are complete and static. We therefore prefer the concept of *credentials* that capture what you know now, evolve by adding new properties over time, and note the confidence in their correctness[9] [31].

Designers use these design spaces in many ways (discussed briefly in the sections that follow), including:
- Comparing and evaluating existing designs (Section 2.1)
- Capturing domain knowledge (Section 2.2)
- Mapping from problem space to solution space (Section 2.3)
- Analyzing quality attributes (Section 2.4)
- Performing tradeoff analysis in a well-understood domain (Section 2.5)

Defining a design space explicitly entails selecting which dimensions to consider out of all the possible properties that require decisions. The goals of the design and intended use for the space should shape this selection, so the design dimensions of interest are highly dependent on context, and they are likely to change as the designer's understanding of the problem evolves. This is a form of Schön's "reflective conversation with the situation" [28].

### 2.1. Comparing and evaluating existing designs

Design spaces can be used to compare existing designs, for critique, for evaluation, for selection among products[10]. This is less subject to the risk of oversimplification than other uses, because the current set of designs is known.

At the Software Designers in Action workshop [41], numerous researchers undertook independent analyses of videos of pairs of designers at a whiteboard, each pair addressing the same design brief for traffic signal simulator. (N.B. An analysis from the workshop using a different lens is in Sec 3.1.) As one of the studies in this workshop, Shaw defined a design space to compare the design decisions made by the three teams, the choices implied by the prompt, and the decisions evident in a commercial product [33]. Figure 2 shows the diversity of choices made. For example, all three groups made different decisions about whether traffic signals should be most closely associated with roads, with intersections, or with an entity that connects roads to intersections; a commercial tool associated the signals with intersections.

---

[1] Experts generate alternatives. #45 Petre and van der Hoek [24] identified 66 insights/practices manifested by expert designers, based on empirical research. We call out relevant insights by their numbers.
[2] Experts solve simpler problems first. #2
[3] Experts keep options open. #30
[4] Experts reshape the problem space. #20
[5] Experts explore different perspectives. #46
[6] Experts design elegant abstractions. #5
[7] Experts prefer simple solutions. #1
[8] Experts invent notations. #28
[9] Experts draw what they need and no more. #26
[10] Experts look around. #14

This representation of the design space emphasizes the dimensions. Following Brooks [4], each major group of decisions is represented as a tree with two kinds of branches: choice and substructure. Substructure branches (not tagged) group independent design decisions; choice branches, flagged with "##", provide alternatives. In some cases, the decision is a numeric value, and the choices are implicit.

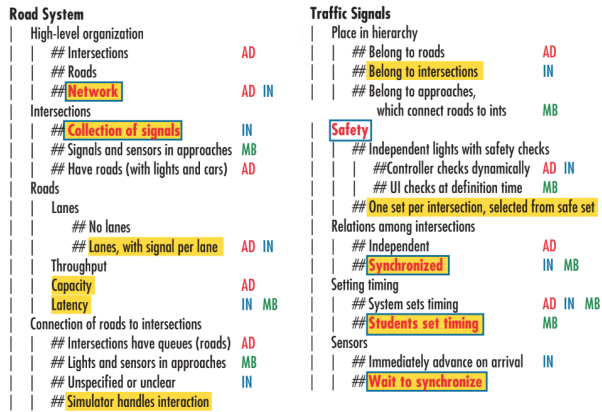

**Figure 2: Part of the comparison of several designs for the traffic signal simulator, showing the decisions implied by the task statement (boxed red text), made by the three teams (*AD*, IN, and *MB*), and made by a commercial product (highlighted in yellow) [32]**

In an example from another discipline, Römer and Mattern created a design space to compare 15 implementations of wireless sensor networks [27]. They observed that the proliferation of such networks with vastly varying requirements and characteristics was making it increasingly difficult to have useful discussions within the community. They identified 14 major dimensions, most with two or more related subdimensions.

Early superscalar hardware required register renaming to resolve performance bottlenecks. Sima studied a decade of register renaming techniques in 26 RISC and 14 CISC commercial superprocessors and identified a hierarchical design space with four major dimensions to help designers understand and explore this complex space [35].

## 2.2. Capturing domain knowledge

Design spaces are sometimes defined to capture and explain knowledge about a domain, especially knowledge that will shape many designs of products[11]. Dimensions of the space are selected to highlight principal distinctions; accordingly, they tend to be fairly static, perhaps evolving as understanding of the domain evolves. However, different aspects of the domain may be significant for different applications, so the dimensions should be selected with that breadth in mind.

In the early development of software architecture styles, Shaw and Clements classified architectural styles in order to establish a uniform descriptive vocabulary, to explain carefully the distinctions among styles, and to lay the groundwork for providing advice about choosing a style appropriate to a problem [34]. The resulting "Boxology" identified six major classes of styles: data flow, call-and-return, interacting processes, data-centered repositories, data-sharing, and hierarchical. Each of these had several specific variants, differing in their constituent parts and their data and control issues. Figure 3 shows a snippet of this space covering one of the major styles.

Following the example of Lane's design space for user interface components [20] (see Section 2.3), the Boxology work identified a few rules of thumb for style selection. For example, it suggested "If a central issue is understanding the data of the application, ... [and] if the data is long-lived, focus on repositories. If the input data is noisy and execution order cannot be predetermined, consider a blackboard." [34].

| Style | Constituent parts | | Control issues | | | Data issues | | | | Control/data interaction | | Type of reasoning |
|---|---|---|---|---|---|---|---|---|---|---|---|---|
| | Components | Connectors | Topology | Synchronicity | Binding time | Topology | Continuity | Mode | Binding time | Isomorphic shapes | Flow directions | |
| **Data-centered repository styles:** Styles dominated by a complex central data store, manipulated by independent computations | | | | | | | | | | | | Data integrity |
| Transactional database [Be90, Sp87] | memory, computations | trans. streams (queries) | star | asynch, opp | w | star | spor lvol | shared, passed | w | possibly | if isomorphic, opposite | ACID[5] properties |
| •Client/server | managers, computations | transaction opns with history[3] | star | asynch. | w, c, r | star | spor lvol | passed | w, c, r | yes | opposite | |
| Blackboard [Ni86] | memory, computations | direct access | star | asynch, opp | w | star | spor lvol | shared, mcast | w | no | n/a | convergence |
| Modern compiler [SG96] | memory, computations | procedure call | star | seq | w | star | spor lvol | shared | w | no | n/a | invariants on parse tree |
| **Key to column entries** | | | | | | | | | | | | |
| Topology | hier (hierarchical), arb (arbitrary), star, linear (one-way), fixed (determined by style) | | | | | | | | | | | |
| Synchronicity | seq (sequential, one thread of control), ls/par (lockstep parallel), synch (synchronous), asynch (asynchronous), opp (opportunistic) | | | | | | | | | | | |
| Binding time | w (write-time--that is, in source code), c (compile-time), i (invocation-time), r (run-time) | | | | | | | | | | | |
| Continuity | spor (sporadic), cont (continuous), hvol (high-volume), lvol (low-volume) | | | | | | | | | | | |
| Mode | shared, passed, bdcast (broadcast), mcast (multicast), ci/co (copy-in/copy-out) | | | | | | | | | | | |

**Figure 3: Snippet of design space for software architecture styles, covering data-centered repositories. Columns correspond to dimensions of the design space [34]**

---

[11] Experts address knowledge deficiencies. #38

## 2.3. Mapping problem space to solution space

If both the desired properties of the design (i.e., the requirements) and the implementation alternatives are described by design spaces (that is, if both a problem space and a solution space are under consideration), it is attractive to look for a mapping that guides the designer from a region of the problem space to promising regions in the solution space[12].

Lane did this analysis for user interface software structures [20]. He interviewed designers of six software systems to discover the characteristics of the user interface components of their systems and the implementation choices they had made. Based on this, he created detailed functional and structural design spaces. Figure 4 shows the principal structure of these spaces; the full functional space has 25 dimensions, each with 3 to 5 alternatives, and the full structural space has 19 dimensions, each with 2 to 7 alternatives. He developed a set of design rules for a recommendation engine that took as input a functional description and produced a ranked set of recommendations for the structural choices; he validated this statistically against the actual designs produced by his subjects. He also produced a set of over three dozen narrative design rules for use by human designers, e.g.: "High user customizability requirements favor external notations for user interface behavior. Implicit and internal notations are usually more difficult to access and more closely coupled to application logic than are external notations."

```
Functional Dimensions                Structural dimensions
1. External requirements             1. Division of functions & knowledge
   a. Application characteristics       a. Application interface
   b. User needs                        b. Device interface
   c. I/O devices                    2. Representation issues
   d. Computer sys environment          a. Means of user interface definition
2. Basic interactive behavior           b. Representation of application info
   a. Interface class                   c. Data reps for communication
   b. Flexibility of interaction        d. Representation of interface state
      sequencing                     3. Control flow, comm, synch issues
3. Practical considerations             a. Control flow
   a. Portability of applications       b. Communication mechanisms
   b. Adaptability of UI system         c. Synchronization issues
```

**Figure 4. Overview of functional and structural design spaces for user interface structures [20]**

Baum et al. proposed an extension of Lane's model as a design aid for software architecture [3]. They added correlations between dimensions in the functional and structural spaces to show strict dependence, incompatibility, or dependencies that create tradeoff decisions[13]. They also expressed design rules between the functional and structural spaces as correlations.

A design space for self-adaptive systems was developed by Brun et al. to guide designing such a system based on given requirements [5]. It identifies five clusters of design decisions related to control aspects of self-adaptive systems and represents them as questions to be answered by the designer.

## 2.4. Analyzing quality attributes

In addition to the largely-descriptive, largely-qualitative examples above, design spaces are also created and explored to analyze or predict quantitative or formal attributes of the designs. This is particularly challenging in modern systems that involve uncertainties arising from lack of control over third-party components, physical components of cyberphysical systems, or human behavior.

Cámara et al. developed a technique for making probabilistic guarantees about such systems [6][7]. Their design space is the set of possible software configurations that are defined by and generated from a formal model that includes structural constraints (architectural style) and application-specific constraints. They explore the design space by further filtering this set of configurations, for example with additional constraints, and quantifying the probabilities of outcomes associated with quality attributes. Unlike most examples in this section, which are represented by explicitly enumerating alternatives, the Cámara design spaces are represented by formal models.

## 2.5. Performing tradeoff analysis in a well-understood domain

Disciplines often share and compare results using a consensus model of a well-understood example. Historically, "stack" served this role for programming languages as did "dining philosophers" for synchronization algorithms and "lift" and "library" for intermediate-level software design. A shared design space can support this sharing with a concrete representation of the choices in a well-understood domain.

In the domain of hardware/software co-design, a class of performance-critical signal-processing problems involves configuring hardware-software allocation and parameterization for system-on-a-chip (SoC) applications. "During system-level design space exploration (DSE), system parameters (e.g., the number and type of processors, and the mapping of application tasks to architectural resources) are considered. The number of design instances that need to be evaluated during such DSE to find good design candidates is large … techniques are needed to optimize the DSE process" [40]. Solutions must balance tradeoffs among chip area, latency, and power [44].

Approaches to achieving designs with better performance include genetic algorithms [37][40], parameter dependency for high-level pruning with genetic algorithms to find a Pareto-optimal tradeoff subspace [21], and semi-random heuristics combined with integer linear programming [44]. Details of the design spaces may differ, but the point of view that performance optimization is coupled to search space exploration is strong. The research is heavy on formalism and includes empirical results.

Kang et al. addressed a similar problem of mapping tasks to devices with a design space exploration technique that systematically eliminated equivalent candidates, based on a user-defined equivalence [19]. This enabled them to generate the relevant candidates from the space in a cost-effective manner.

## 3. Integrative Design Spaces

Some design processes refer to working in a design space, but don't represent the rich and complex space of alternatives explicitly. They often are more concerned with the design process that exposes and contrasts alternatives, than with the enumeration or representation of the alternatives per se. Hence Woodbury and Burrow [43] argue that design spaces and their representations are helpful—maybe even necessary—tools, but that they should not dictate the design, which requires reasoning beyond representation details[14]. Integrative designs are concerned with a broad exploration of the design space(s), with a focus on identifying the key design considerations—and with ongoing attention to the interactions among the design requirements and features, and the dependencies among the choices explored. Hence, the exploration prioritizes the "essence" of the problem and defers other elements[15]. Schön characterized this integrative approach [28]: "Let us search ... for an epistemology of practice implicit in the artistic, intuitive processes which some practitioners bring to situations of uncertainty, instability, uniqueness, and value conflict".[16]

---

[12] Experts draw the problem as much as they draw the solution. #25

[13] Experts make tradeoffs. #34

### 3.1. Co-evolution of problem & solution spaces

Dorst and Cross studied nine professional designers working on a design for a litter disposal system for trains [14]. They found that the designers start by exploring the problem space and positing a partial solution[17]. They then explore the partial solution in the solution space, which provides insights about the problem space as in Figure 5. Rinse, lather, repeat.

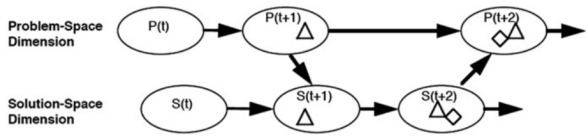

**Figure 5. Problem-solution space co-evolution [14]**

Curtis reported similar design behavior in software designers [10]. Guindon [17] found that designers bounced among levels of abstraction and developed mental models of the problem domain and the solution concurrently[18]. Figure 6 shows this, with Curtis' overlay showing the expected attention of a waterfall model. This opportunistic behavior occurred even when a trained software developer attempted to do top-down development. Curtis observed that design insights reorganize the designer's model of the problem domain, which creates new relations between the problem and solution domains[19].

In one of the studies from the Software Designers in Action workshop [41], Tang et al. analyzed the effectiveness of the two teams' design strategies [39]]. (N.B. An analysis from the workshop using a different lens is in Sec 2.1.) One of these strategies is problem-solution co-evolution, in which both the problem space and the solution space are developed and refined together; they associate this with the process of Figure 5.

### 3.2 Tightly-coupled decisions

Penders et al. [23] describe an exploration-based design strategy that identifies different design-space dimensions and uses inter-dependency rules between those dimensions to make systematic design decisions. "Inter-dependency rules describe the influence design choices in one dimension have on other design dimensions…" and "Inter-dependency rules indicate that, when limiting the choices in one dimension, there is a direct effect on possible design choices in its dependent dimension(s). Thus, inter-dependencies form a navigation guide for the designer, as they imply correlations between dimensions: after inspecting a design dimension a logical next step is to inspect dependent dimensions." [23]

Hence, the interactions between decisions are a principal focus of the design space analysis. Penders et al. argue that "One of the important advantages of the presented exploration-based design strategy is its ability to enforce a systematic decision for each dimension in the space" [23]. Figure 7 shows such inter-dependency rules in the context of hardware/software co-design.

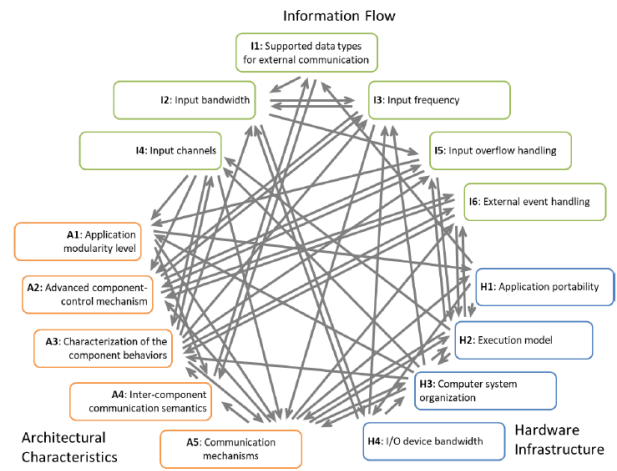

**Figure 7. Dependency relationships among decisions [23]**

The richness of the interactions among decisions is well illustrated by an example outside of software engineering, the selection of resins for molding plastic parts [18]. Requirements for a molded part might include stiffness, dimensional stability, and resistance to UV; production requirements might include cost of resin, complexity of molding process, and reliability of supplier. "Resin properties influence part performance. … A gain in one property often coincides with a loss in another … Every resin property required for an application influences material selection in varying degrees."

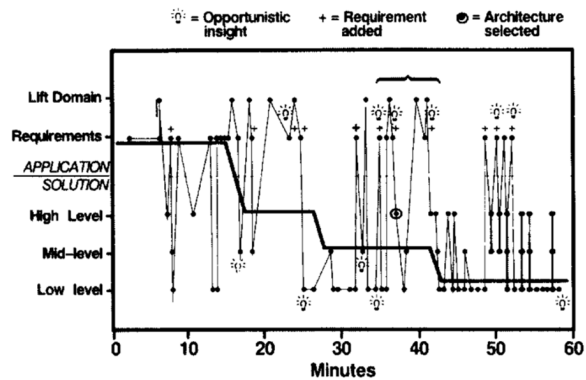

**Figure 6. Process map of a software design [17], overlay by [10]**

---

[14] Experts think about what they are not designing. #60
[15] Experts focus on the essence. #37
[16] Experts adjust to the degree of uncertainty present. #36
[17] Experts make provisional decisions. #31
[18] Experts move among levels of abstraction. #47
[19] Experts re-assess the landscape. #61

## 3.3 Implicit design spaces with coupled decisions

Sometimes the important design decisions are related to keeping a system in balance, with an appropriate balance of resources. Here the explicit description of the design space may take a back seat to reasoning about the interactions, so the design space is implicit in the problem analysis[20]. Devereaux [11] considers such a case of logistic considerations for swift operational army maneuvers in the pre-gunpowder period, when "march fast" was the secret to major victories. The essential question is the relation between distance, speed, and force size for moving an army through enemy territory. Devereaux' analysis goes through these steps:

- The limiting factor is meeting nutritional demands; adult men need ~3000 calories/day [identified key factor: food]
- Historically, this has worked out to ~3 lb of food per person per day, assuming water is available [new parameter: weight per person per day]
- Marching loads are ~80-100 lbs per person with non-food load accounting for all but ~30lb, so ~10 days rations [new factor: food transport]
- The duration can be extended by carrying additional supplies. In addition to food for soldiers, you need food for porters, or horses, or mules. In practice this can only extend the duration to ~40 days [relax the 10-day limit on food transport with a pack train, but at cost of more food and less speed]
- More supplies can be acquired en route by foraging, but this has limited capacity and slows the movement; there is a tradeoff between short operational range at good speed and slower movement with unwieldy supply lines [new parameter for food transport, with associated costs]
- The army will have many non-soldiers performing various duties; this can range from 20% to nearly 100% more human mouths to feed, plus animals [discovered a new parameter, the "tooth to tail" ratio, the ratio of soldiers to support people]
- Bigger armies are slower and harder to coordinate and secure [size affects speed]

The analysis goes on in considerably more detail, but this summary already shows a design space with factors including food, weight of food per day, food transport, ratio of soldiers to noncombatants, and number of horses/mules; these are major determinants of size of army, rate of movement, duration in days, and operational range. These factors interact in ways that sharply constrain the feasible points in the design space.

## 4. How Designers Use Design Spaces

As discussed, design spaces embody the design alternatives for problem domains or applications, providing different lenses on the design possibilities[21]. What emerges from the sample of design spaces is their role as tools for designers, assisting them:

- To systematize understanding of the alternatives (scope and characteristics) (Section 4.1)
- To systematize navigation of the alternatives (exploration and decision processes) (Section 4.2)
- To support orderly evolution of understanding (through the process of exploration) (Section 4.3)
- To support the orderly reuse of prior art (when existing solutions are appropriate) (Section 4.4).

## 4.1 Systematize understanding of alternatives

The examples demonstrate the utility of various design spaces in helping to identify key design considerations, alternatives, and the interactions between them. These may be known—or they may emerge as part of the design space exploration, or in the contrast between different design spaces. Hence, by helping designers externalize (and hence examine) their understanding, design spaces help designers to check for coverage by prompting them to consider alternatives systematically within the space. On the other hand, the systematic exploration can help identify regions of the design space that are not currently relevant, feasible, or prioritized—helping to reduce the alternatives in a reasoned way.

Schön [29] discussed this as problem framing: "In order to formulate a design problem to be solved, the designer must frame a problematic design situation: set its boundaries, select particular things and relations for attention, and impose on the situation a coherence that guides subsequent moves."

## 4.2 Systematize navigation of the alternatives

In order to achieve this understanding, designers need to navigate the design space effectively and systematically. Many authors (e.g., Simon, Newell, Brooks) characterize design as search in a design space. Whitworth and Ahmad further characterize that search as choosing a point in the design space: "System design is choosing a point in a multi-dimensional design criterion space with many "best" points, e.g. a cheap, heavy but powerful vacuum cleaner, or a light, expensive and powerful one. The efficient frontier is a set of "best" combinations in a design space. Advanced system performance is not a one-dimensional ladder to excellence, but a station with many trains to many destinations." [42]

That navigation—how designers explore the design space, and how they choose among alternatives—can take different forms, influenced not only by the design objectives but also, for example, by their domains, experience, representations, processes, etc. Computer scientists are naturally inclined to view this search through an algorithmic lens, expecting systematic, orderly, potentially exhaustive traversal of the search space. Brooks [4] notes the appeal of a rational model that systematically searches the design space but also observes that this model is vastly oversimplified: real designers just don't work that way[22]. In particular, designers often satisfice, stopping the search when an acceptable, if not optimal solution is found[23]. Guindon [17] found an opportunistic mix of breadth-first and depth-first search (see Figure 6). Brooks reports critiques by Cross [9], Schön [28], Akin [1], Royce [26] and others that show not only complex navigation of the search space but also evolution of the space itself as design proceeds. The aspiration to systematic and rational reasoning about design, and the pragmatics of working within the constraints of human cognition are the essence of design space exploration; Dorst [12] still reported in 2006 that the rational, systematic model still looms large in the field.

---

[20] Experts retain their orientation. #59
[21] Experts change notation deliberately. #49
[22] Experts use design methods (selectively). #17
[23] Experts know when to stop. #42

The representation of the design space may also influence the way it's navigated. A hierarchical representation in which each decision shapes the alternatives available for future decisions tends to encourage making decisions in the order embedded in the hierarchy. "What remains unclear is to what extent such strategies are conditioned by the external memory aids available to designers" [43].

### 4.3 Support orderly evolution of understanding

Designers' understanding of the design space evolves through the process of exploration. As Cross [8] reported: Schön [28] also pointed out that "the work of framing is seldom done in one burst at the beginning of a design process." This was confirmed in Goel and Pirolli's [15] protocol studies of several types of designers (architects, engineers, instructional designers). They found that "problem structuring" activities not only dominated at the start of the design task, but also re-occurred periodically throughout the task[24]. This points to the interaction between the problem space and the solution space.

As the understanding of the design space evolves, the process may reveal critical factors (or change the priorities of different factors) and/or reveal assumptions [25]. As a result, new possible solutions may emerge. Some of the examples highlight the contrasts between design spaces, and in particular the interchange between the problem space and the solution space. "Many studies of expert design behavior suggest that designers move rapidly to early solution conjectures, and use these conjectures as a way of exploring and defining problem-and-solution together"[26] [8].

Dorst asserts, "We urgently need a strong descriptive and analytical framework to help us understand what is actually going on in the "upwards jump" from solution to problem, and how we can safeguard against the misuse and manipulation of emergence" [13].

### 4.4 Support orderly reuse of prior art

Although this discussion has prioritized innovative or radical design, design spaces play a key role in routine or normal design as well, where for each instance the task is selecting an established design from a well-defined family[27].

In typography, a designspace [sic] is an element of a typeface design that controls how a variable font's appearance changes (by interpolation) as its variation axes are adjusted [16][38]. A variable font with weight and width axes has a 2-dimensional space, as in Figure 8; some parametric fonts have 10 or more dimensions of variability. This allows users of the font to tune, but not to otherwise redesign, the font. Font designers, of course, use the designspace to represent the range of variability they intend [38].

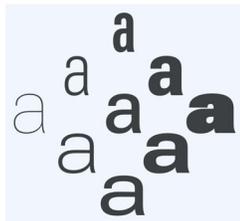

**Figure 8. Points in a font design-space [16]**

In the domain of civil engineering, Pennsylvania mandates the use of the Bridge Automated Design and Drafting (BRADD) software for new and replacement single span bridges [22]. This software automates the bridge design process from problem definition through contract drawings for specific types of simple single-span bridges. The scope is concrete, steel, and prestressed concrete bridges with spans from 18 to 200 ft, with a variety of site-specific parameters. It offers 5 types of superstructures and 3 types of abutments. It consists of 4600 Fortran routines (730,000 lines of code), 590 dimensions/parameter/data files (92,500 lines of data), and 260 graphic details containing 976 overlays, and a Visual Basic user interface of 83,000 lines of code.

### 5. Conclusion

We sampled the literature, primarily in software engineering and principally in system-level design, to find examples of how designers use design spaces. This is not a comprehensive survey, let alone a systematic review. We reached for the low-hanging fruit, a set of examples rich enough to show the diversity of uses in practice. We included examples from other fields to emphasize that design spaces are integral to design generally, not specifically a software construct. This serves our main purposes: to raise the visibility of this design tool, to improve communication by clarifying its various meanings, and to exhibit multiple uses that may stimulate further development.

This sampler shows the wide diversity in content, representation, and use. The main point is the idea of structuring knowledge about a design; the specifics are tools that help designers think, not rigid rules that require adherence[28].

You, the reader, may use "design space" in yet another way. That's fine, and some future extension of this work might include your use. You should, however, be aware of other ways that other people use the term so that you and they can communicate without confusion; the other examples may inspire you to extend your own.

The designer's principal responsibility is to understand the client's needs and commit to finding a solution that satisfies those needs. All of the models, methods, notations, representations, frameworks, and processes of software engineering—including design spaces—are tools to that end, not ends in themselves. The designer should select the ones that help with the problem at hand. These examples show that design spaces deserve their place in this toolkit, with details of their role in any particular design subject to the judgment of the designer.

### ACKNOWLEDGMENTS

We thank our colleagues who have contributed to this sampler through their research and discussion. This work was partially supported by the Alan J. Perlis chair of Computer Science at Carnegie Mellon University.

---

[24] Experts design throughout the creation of software. #22
[25] Experts are alert to evidence that challenges their theory. #57
[26] Experts make provisional decisions. #31
[27] Experts prefer solutions that they know work.#13
[28] Experts do not feel obliged to use things as intended. #23
  Experts use notations as lenses, rather than straightjackets. #21